\documentstyle[aps,prl,rotate]{revtex}

\input{psfig}
\input{epsf}

\pssilent

\begin{document}

\title{Search for Neutral Heavy Leptons in a High-Energy Neutrino Beam }
\author{A.~VAITAITIS$^{2}$, R.~B.~DRUCKER$^{6}$, J.~FORMAGGIO$^2$, 
T.~ADAMS$^{4}$, A.~ALTON$^{4}$,
S.~AVVAKUMOV$^{7}$, L.~de~BARBARO$^{5}$, P.~de~BARBARO$^{7}$, R.~H.~BERNSTEIN%
$^{3}$, A.~BODEK$^{7}$, T.~BOLTON$^{4}$, J.~BRAU$^{6}$, D.~BUCHHOLZ$^{5}$,
H.~BUDD$^{7}$, L.~BUGEL$^{3}$, J.~CONRAD$^{2}$, R.~FREY$^{6}$, 
J.~GOLDMAN$^{4}$, M.~GONCHAROV$^{4}$, D.~A.~HARRIS$^{7}$,
R.~A.~JOHNSON$^{1}$, S.~KOUTSOLIOTAS$^{2}$, J.~H.~KIM$^{2}$, M.~J.~LAMM$^{3}$%
, W.~MARSH$^{3}$, D.~MASON$^{6}$, C.~McNULTY$^{2}$, K.~S.~McFARLAND$^{3,7}$,
J.~MONROE$^{2}$,
D.~NAPLES$^{4}$, P.~NIENABER$^{3}$, A.~ROMOSAN$^{2}$, W.~K.~SAKUMOTO$^{7}$,
H.~SCHELLMAN$^{5}$, M.~H.~SHAEVITZ$^{2}$, P.~SPENTZOURIS$^{2}$, E.~G.~STERN$%
^{2}$, B.~TAMMINGA$^{2}$, M.~VAKILI$^{1}$, V.~WU$^{1}$, U.~K.~YANG$^{7}$,
J.~YU$^{3}$, G.~P.~ZELLER$^{5}$, and E.~D.~ZIMMERMAN$^{2}$}
\date{\today}
\address{
$^1$University of Cincinnati, Cincinnati, Ohio 45221 \\
$^2$Columbia University, New York, New York 10027 \\
$^3$Fermi National Accelerator Laboratory, Batavia, Illinois 60510 \\
$^4$Kansas State University, Manhattan, Kansas 66506 \\
$^5$Northwestern University, Evanston, Illinois 60208 \\
$^6$University of Oregon, Eugene, Oregon 97403 \\
$^7$University of Rochester, Rochester, New York 14627 \\
}
\normalsize

\maketitle

\begin{abstract}
A search for neutral heavy leptons (NHLs) has been performed using an
instrumented decay channel at the NuTeV (E-815) experiment at Fermilab. The
decay channel was composed of helium bags interspersed with drift chambers,
and was used in conjunction with the NuTeV neutrino detector to search for
NHL decays. The data were examined for NHLs decaying into muonic final
states ($\mu \mu \nu $, $\mu e\nu $, $\mu \pi $, and $\mu \rho $); no
evidence has been found for NHLs in the 0.25 - 2.0 GeV mass range. This analysis
places limits on the mixing of NHLs with standard light neutrinos at a level
up to an order of magnitude more restrictive than previous search
limits in this mass range. 

\vspace{0.1in} 
\end{abstract}

%\twocolumn[\maketitle

\newpage 
\narrowtext
\twocolumn

%\pssilent

\begin{figure}[tbh]
{\Large \bf a)}
\centerline{\psfig{figure=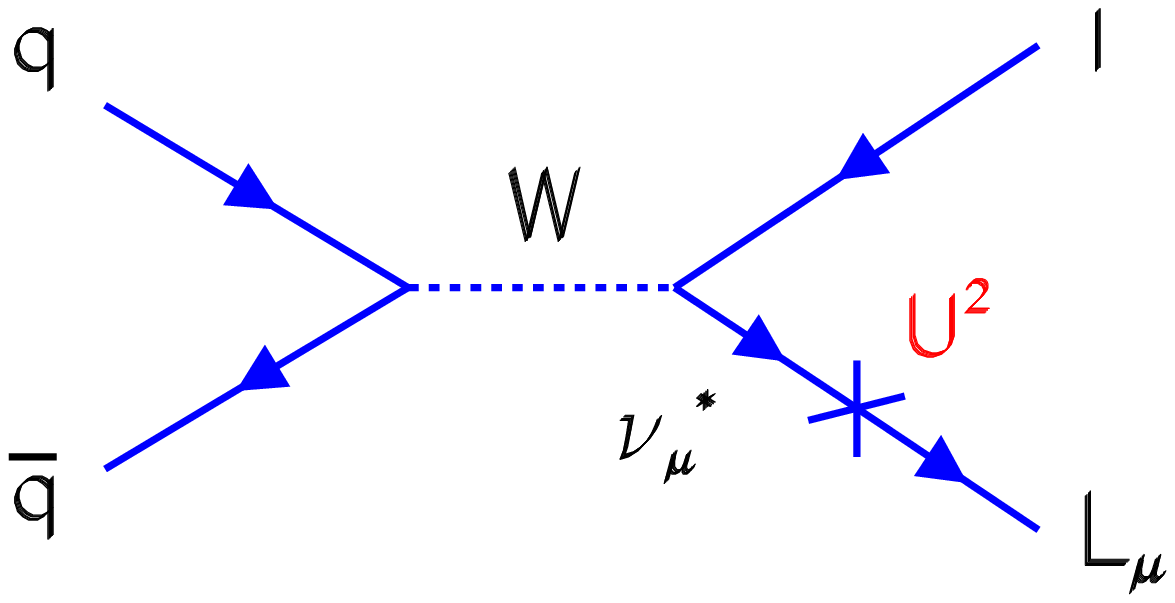,width=0.3%
\textwidth,bbllx=66pt,bblly=262pt,bburx=410pt,bbury=443pt}} %

{\Large \bf b)}
\centerline{\psfig{figure=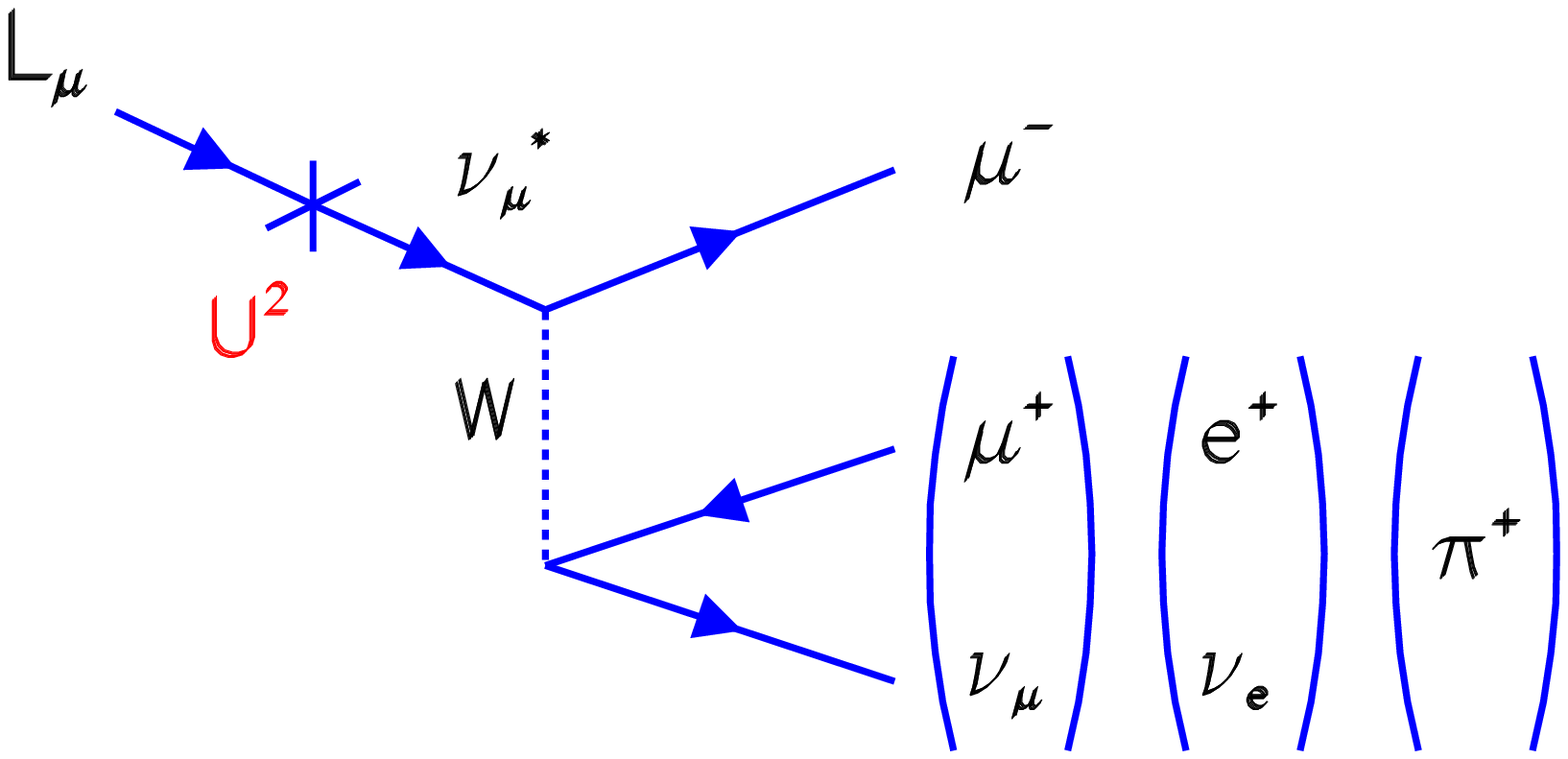,width=0.3%
\textwidth,bbllx=25pt,bblly=200pt,bburx=508pt,bbury=442pt}}
\caption{Feynman diagrams for the (a) production and (b) decay of neutral
heavy leptons ($L_{\protect\mu }$). $|U|^{2}$ is the mixing 
between the muon neutrino and the NHL. (The alternative allowed decay via a
Z boson is not shown.)}
\label{feynman}
\end{figure}

Various extensions\cite{grl,pdg} to the Standard Model predict neutral heavy
leptons (NHLs) which can mix with the standard light neutrinos. In these
extensions, the NHLs are weak iso\-singlets that do not couple directly to
the $Z$ and $W$ bosons, but can decay via mixing with the Standard Model
neutrinos. Figure ~\ref{feynman} shows one possible set of
tree-level diagrams for NHL production and decay. The NuTeV (E815) neutrino
experiment at Fermilab has made a sensitive search for these NHLs by
combining the capabilities of a high intensity neutrino source with an
instrumented decay region. 

In these extended models\cite{grl},
 the NHL lifetime depends on the mixing parameter $|U|^{2}$ and the 
mass of the NHL. They are expected to decay (e.g.~Fig.~\ref{feynman}b) 
into a neutrino and two charged leptons, into a lepton and two
quarks, or into three neutrinos. 

NHLs may be created in the NuTeV beamline by the decays of secondary mesons
produced by the Tevatron proton beam. During the 1996-1997 fixed-target run at
Fermilab, NuTeV received $2.54\times 10^{18}$ 800~GeV protons on a beryllium
oxide production target with the detector configured for this search. A 
sign-selected quadrupole train focused 
secondary $\pi $ and $K$ mesons down a beamline 7.8~mrad from the primary
proton beam direction. $1.13\times 10^{18}$ protons were received with
the magnets set to focus positive mesons, and $1.41\times 10^{18}$ protons
with negative meson focusing. The mesons could 
decay into NHLs as shown in Fig.~\ref{feynman}a. The 
production of secondary pions and kaons
was simulated using the parameterization in \cite{malensek}; the Decay Turtle 
program~\cite{turtle} simulated the propagation of charged particles through 
the beamline. 

NHLs may also be produced by prompt decays of charmed mesons produced by 
incident protons on the BeO target and proton dumps.
These processes were simulated using a Monte Carlo program based on measured
production cross sections\cite{charm}. The effects of decay phase space,
NHL polarization, and helicity suppression\cite{joe} were included in the
simulation of the production and decays. %Figure~\ref{fig:nhlp} shows
%examples of the momentum distribution of NHLs produced in the NuTeV
%beamline. 
For NHLs of mass 1.45~GeV from D meson decay, the average momentum
was $\sim $100~GeV; for the 0.35~GeV NHLs coming mainly from K decay, the
average momentum was $\sim $140~GeV. 

This analysis reports the results of a search for NHLs with masses between
0.25 to 2.0 GeV, which decay with a muon in the final state. The primary NHL
decay modes of this type are $\mu e\nu ,\mu \mu \nu ,\mu \pi ,$ and $\mu
\rho $. In the standard mixing model\cite{grl,tim}, the expected ratio of
decay rates for $\mu \pi $ : $\mu e\nu $ : $\mu \mu \nu $ is 3.5 : 1.6 :
1.0. The $\mu \rho $ channel is significant only for NHL masses above 1~GeV,
where it is comparable to the $\mu \pi $ mode. 

\begin{figure}[t]
\centerline{\psfig{figure=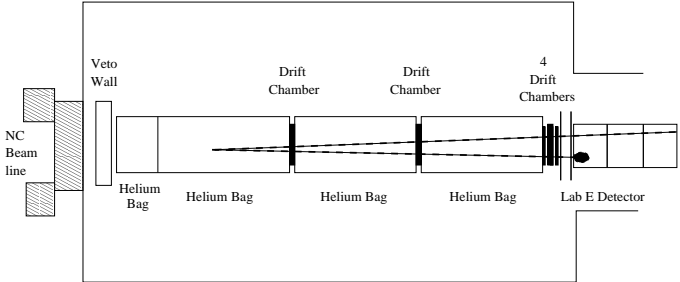,width=0.5\textwidth}}
\caption{A schematic diagram of the NuTeV decay channel. The beam enters
from the left; at the far right is the Lab E calorimeter. An example NHL decay
to $\protect\mu \protect\pi $ is also shown. The event appears as two tracks
in the decay channel, and a long muon track and a hadronic shower in the
calorimeter.}
\label{fig:dkchannel}
\end{figure}

In order to detect NHL decays, an instrumented decay region (the
``decay channel'') was constructed 1.4 km downstream of the production target.
The decay channel (Fig.~\ref{fig:dkchannel}) was designed to identify
signatures characteristic of NHLs: a neutral particle decaying to two
charged particles within the decay channel. A 4.6~m~$\times$~4.6~m 
array of plastic
scintillators vetoed charged particles entering from upstream. The decay channel
was 34 m long, interspersed with 3~m~$\times$~3~m multi-wire argon-ethane
drift chambers positioned at 15.4~m, 25.2~m, 35.5~m, 36.8~m, and 38.5~m 
downstream of the veto array. Tracks were
reconstructed from drift chamber hits and grouped to form candidate decay
vertices. Typically, a vertex from a NHL of mass 1.15 GeV was reconstructed
with a resolution of 0.04 cm in the transverse direction and 28 cm
in the longitudinal direction. The region between the drift chambers 
was filled with helium
contained in cylindrical plastic bags 4.6 m in diameter. By displacing the
air with helium, the number of background neutrino interactions between the
chambers was reduced by a factor of 7. 
%The region between the chambers was filled with helium, which is
%$\times$ 7.2 less massive than air, contained
%in 4.6~m diameter plastic bags. 

The Lab E neutrino detector\cite{detector}, located immediately downstream
of the decay channel, provided final state particle identification,
energy measurement, and triggering. This detector consisted of a 690-ton 
iron-scintillator
sampling calorimeter followed by a toroidal muon spectrometer. 
Drift chambers were positioned every 20~cm along the length of
the calorimeter, and 2.5 cm-thick liquid scintillator counters were
interleaved with 10~cm steel plates along the length of the calorimeter. 
The trigger selected events with a penetrating muon or at least 6~GeV of
hadronic or electromagnetic energy in the calorimeter. 
Offline, hits in the calorimeter drift chambers 
were analyzed to determine if they formed a track (consistent with a muon)
or a cluster of hits (consistent with an electron or pion shower). 

Tracks found in the decay channel were linked to clusters or tracks in the 
calorimeter. Pulse
height information from the scintillation counters was used to 
determine hadronic or electromagnetic energy
deposition; the hadronic energy resolution of the calorimeter was $\frac{%
\sigma }{E}=0.024 \oplus \frac{0.87}{\sqrt{E}}$, and the
electromagnetic energy resolution was $\frac{\sigma }{E}=0.040 \oplus \frac{%
0.52}{\sqrt{E}}$. Muon energies
were determined either by the toroid spectrometer (with 11\% resolution) or
by range (with a resolution of 155 MeV). The energy of muons which exited
through the side of the calorimeter was determined from the track's multiple
scattering, with a resolution of 25\% for energies lower than 85 GeV. 

The following processes were the main backgrounds to this search: 1)
neutrino interactions within the helium; 2) neutrino 
interactions in the
material upstream of the decay channel, where a neutral kaon survived to the
decay channel and decayed in the fiducial volume; and 3) neutrino interactions in the
material surrounding the decay channel. Neutrino interaction
backgrounds could arise from low multiplicity resonance production or from
deep inelastic scattering (DIS). Resonance production was simulated
using the calculations of Belusevic and Rein \cite{lownu}; such events
were  characterized by a high-energy muon accompanied by a low-energy
pion track.  The Lund Monte Carlo program was used to simulate DIS 
events \cite {lund}. While these were generally of high
multiplicity,  events with few or 
poorly-reconstructed tracks could contribute to the two-track background.
The detector was simulated using the GEANT~\cite{geant} Monte Carlo, which
produced a hit-level simulation of raw data including beam and cosmic ray
related noise hits. These were processed using the same analysis routines as
those used for the data. 

Backgrounds from cosmic rays were estimated using
a sample of cosmic ray muons which interacted in the target calorimeter; 
an upper limit of $10^{-3}$ background events was determined. We
have therefore ignored cosmic rays in the final background estimate.

%Cosmic ray and other beam-unrelated
%backgrounds were determined by analyzing a sample of events collected during
%beam-off gates throughout the run. 
Event selection criteria maintained high efficiency for the NHL signal while
minimizing known backgrounds. Cuts fell into two broad categories:
reconstruction and kinematic. 

Reconstruction cuts isolated events with a two-track decay vertex within the 
decay channel fiducial volume and no charged particle identified in the
upstream veto system. The two tracks were required to be well-reconstructed, 
have an accompanying energy measurement, and form a
common vertex. Large angle tracks arising from cosmic rays were removed 
by requiring tracks to form an angle of less than 
0.1 radians with the beam direction. Exactly two
tracks were required, both projecting to the calorimeter, with
at least one of the two identified as a muon. To ensure good particle
identification and energy measurement, muons were required to have an energy
greater than 2.0~GeV; an energy greater than 10.0~GeV was required for
electrons or hadrons. (The latter cut eliminated the low-energy pions
associated with resonance production.) The reconstructed two-track decay
vertex was required to be at least 1 m from any material in the drift
chambers. 

The kinematic cuts were designed to remove the remaining DIS and resonance
backgrounds. The effective scaling variables $x_{{\rm eff}}$ and $% 
W_{{\rm eff}}$ were calculated for each event under the following
assumptions: 1) the event was a neutrino charged current interaction ($\nu
N\rightarrow \mu N^{\prime }X$); 2) the highest energy identified muon 
was the outgoing particle from the
lepton vertex; and 3) the missing transverse momentum in the event was
carried by an undetected final state nucleon. Specifically, $x_{{\rm eff}}\equiv\frac{%
Q_{\rm vis}^{2}}{2m_{p}\nu _{vis}}$ and $W_{\rm eff}\equiv\sqrt{m_{p}^{2}+2m_{p}\nu
_{\rm vis}-Q_{\rm vis}^{2}}$, where $Q_{\rm vis}^{2}$ is the reconstructed 
momentum
transfer squared, $m_{p}$ is the mass of the proton, and $\nu _{\rm vis}$ is 
the reconstructed hadron energy. Requiring $x_{{\rm eff}}<0.1$ and 
reduced backgrounds from DIS; requiring $W_{\rm eff}>2.0$~GeV removed
quasielestic and resonance backgrounds. 
Since we could not reconstruct the true mass of the NHL due to the missing 
neutrino, a cut was applied on the ``transverse mass,''
$m_{T}\equiv|p_{T}|+\sqrt{p_{T}^{2}+m_{V}^{2}}$. $p_{T}$ is the 
component of the total momentum perpendicular to the beam direction, 
and $m_{V}$ is the invariant mass for the two charged tracks. Requiring
$m_T < 3.0$~GeV removed additional DIS background.

The NHL and background acceptances and reconstruction efficiencies were
calculated from the hit-level Monte Carlo simulation. For a range of NHL
masses from 0.25 to 2.0 GeV, mesons were allowed to decay in the beamline to
muons and NHLs. The NHLs from these decays were propagated to the decay
channel, weighted by their decay and the meson production probability, and
allowed to decay in the decay channel producing hits in the drift chambers
and energy deposition in the Lab E calorimeter scintillation counters. These
simulated raw data were reconstructed with the same programs used for the
data including the above reconstruction and kinematic cuts. For NHL decays
generated within the decay channel fiducial volume, the average
acceptance was 23\% for masses between
0.25 and 2.0 GeV. From the background simulation, the calculated background
contributions for the full data sample after cuts are listed in Table~\ref
{bkgnd}. 

\begin{figure}[bt]
\centerline{\psfig{figure=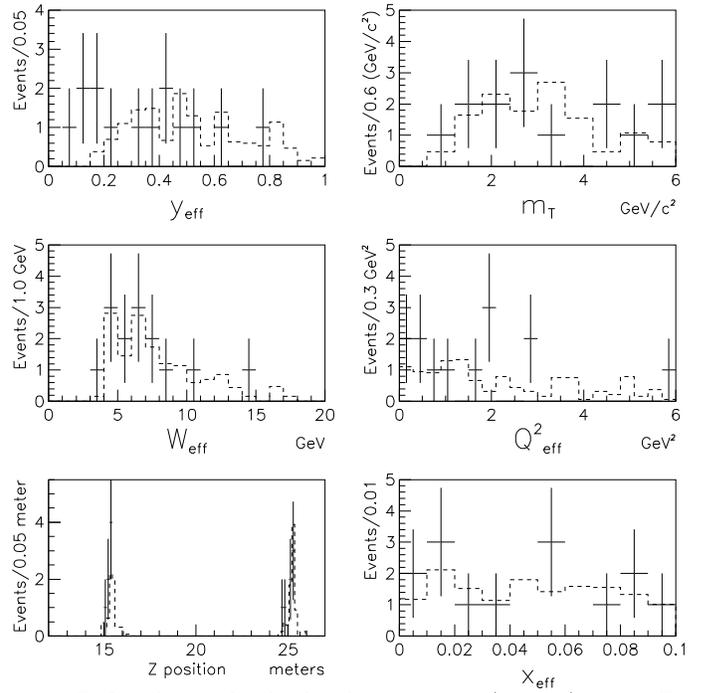,width=0.5\textwidth}}
\caption{Kinematic distributions for data (crosses) and MC DIS background 
(dashed) events with two-track vertices reconstructed in the decay channel
drift chambers. MC events are absolutely normalized to the number of
protons on target. The DIS reconstructed
variables $y_{{\rm eff}}$, $W_{{\rm eff}}$, $Q^2_{{\rm eff}}$,
$x_{{\rm eff}}$ and $m_T$ are defined in the text.
$Z$ positions are referenced to the veto array; spikes in the $Z$
distribution correspond to the locations of the decay channel drift
chambers.
}
\label{fig:chamber}
\end{figure}

Neutrino events in the drift chambers and DIS ($>$2 track) events
in the helium have been used to cross check the Monte Carlo calculations of
backgrounds and efficiencies. For example, as a check on the two-track
reconstuction efficiency, we have compared the fourteen two-track data
events with vertices in the drift chambers to the Monte Carlo prediction of
15 $\pm $ 2 events. Figure~\ref{fig:chamber} shows comparisons of
distributions for the
previously introduced DIS variables, $x_{{\rm eff}}$, $W_{{\rm eff}}$,
$Q^2_{{\rm eff}}$ as well as $m_{T}$ and $y_{{\rm eff}} \equiv
\frac{E_{\nu} - E_{\mu}}{E_{\nu}}$.
Data events with $>2$ tracks also showed good agreement with the Monte Carlo
prediction: 280 events were predicted, and 275 were observed.

For the full NuTeV data sample, no data events passed all cuts. This is
consistent with the expected background of $0.57\pm 0.15$ events. We have
set limits from this null result by using the Monte Carlo prediction for $%
N_{\rm pred}(m_{\rm NHL},|U|^{2})$, the normalized number of NHL events 
expected in
the decay channel as a function of mass and $|U|^{2}$. The 90\% confidence
level limit for a given mass was calculated for the null observation by
finding the $|U|^{2}$ value for each mass such that $%
N_{\rm pred}(m_{\rm NHL},|U|^{2})=2.3$ events. 

The statistical 90\% confidence level limit was modified by the addition of
systematic uncertainties on the number of expected NHLs. These uncertainties
are summarized in Table~\ref{tab:syst}. Experimental uncertainties on the $D$
production cross section \cite{charm} and\ on the $K$ meson flux Monte Carlo 
\cite{turtle} were the dominant contributions to the systematic error. The
systematic uncertainties were incorporated into the 90\% confidence level
limit by adding in quadrature a fractional error term corresponding to the
systematic uncertainty for a given mass. Since the NHL rate was proportional
to $|U|^{4}$, adding the systematic uncertainties increased the $|U|^{2}$
limit by 4\%(14\%) at 0.35(1.45) GeV. 

\begin{figure}[tbp]
\begin{minipage}[c]{\textwidth}                 
\hspace{-0.022\textwidth}\rotate[l]{\psfig{figure=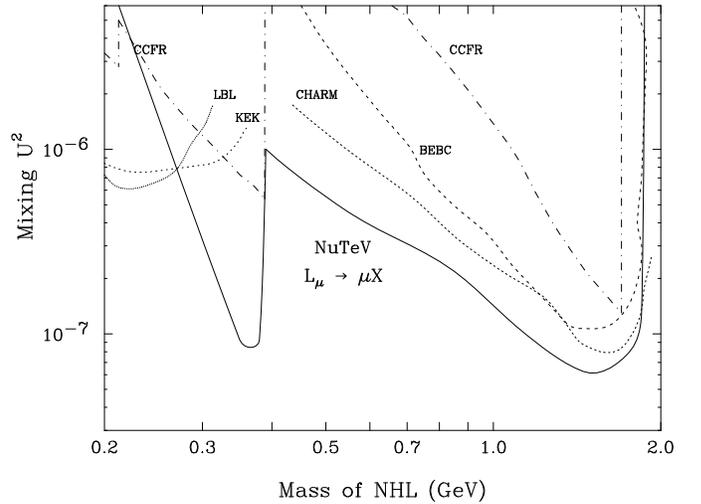
,width=0.37\textwidth}} 
\end{minipage}                                                        
\caption{NuTeV 90\% confidence level limit for U$_{\protect\mu }^{2}$, the
mixing of NHLs to Standard Model left-handed muon neutrinos, as a function
of NHL mass. The solid line in the figure corresponds to the limit with
systematic errors included. }
\label{fig:limits}
\end{figure}

Figure~\ref{fig:limits} shows the limits obtained from this search for the
NHL-$\nu _{\mu }$ mixing parameter $|U^{2}|$ as a function of the mass of 
the NHL.
The limits are for NHL decay modes containing a muon in the standard mixing
model presented above. Also shown in Fig.~\ref{fig:limits} are the results
of previous experiments. \cite
{prev_ccfr,prev_bebc,prev_charm,prev_kek,prev_lbl} 

In summary, we have made a sensitive search for NHLs in the mass range from 
0.25 to 2.0 GeV that mix with muon neutrinos. No evidence for NHL
production was observed. New limits have been set that are up to an order of
magnitude better than previous searches in the lower mass range. 

This research was supported by the U.S. Department of Energy and the
National Science Foundation. We thank the staff of FNAL for their
contributions to the construction and support of this experiment during the
1996-97 fixed target run.

\newpage
\begin{table}[tbh]
\centering
\caption{Backgrounds to the NHL search.}
\begin{tabular}{|r|c|}
%\hline
background source description & number of events \\ \hline
$\nu$ interactions in the helium & $0.56 \pm 0.15$ \\ 
$K^0$ punch-through & $0.005 \pm 0.001$ \\ 
$\nu$ interactions in the material & $0.002 \pm 0.001$ \\ 
%\hline
\end{tabular}
\label{bkgnd}
\end{table}

\begin{table}[tbh]
\centering
\caption{Systematic uncertainties on the sensitivity of the NHL search.}
\begin{tabular}{|r|c|c|c|}
%\hline
& \multicolumn{2}{c}{NHL mass (GeV/$c^2$)}& \\ \cline{2-4}
Source & 0.38 & 0.85 & 1.45 \\ \hline
{D production} & - & {44.6\%} & {44.0\%} \\ 
{K production} & {\ 20.7\%} & - & - \\ 
{$D_S$ production} & - & {5.6\%} & {5.4\%} \\ \hline
{alignment } & {1.0\%} & {0.4\%} & {0.02\%} \\ 
{resolution model} & {9.3\%} & {7.0\%} & {0.8\%} \\ 
{reconstruction eff.} & {17.0\%} & {19.1\%} & {17.0\%} \\ \hline
{\bf TOTAL SYST.} & {28.4\%} & {49.3\%} & {47.5\%} \\ 
%\hline
\end{tabular}
\label{tab:syst}
\end{table}

\newpage

%\begin{figure}[tbh]
%\centerline{\psfig{figure=nhl_prod.eps,width=0.2%
%\textwidth,bbllx=66pt,bblly=262pt,bburx=410pt,bbury=443pt}} %
%\centerline{\psfig{figure=nhl_w.eps,width=0.3%
%\textwidth,bbllx=25pt,bblly=200pt,bburx=508pt,bbury=442pt}}
%\caption{
%Feynman diagrams for the (a) production and
%(b) decay of neutral heavy leptons ($L_\mu$).  
%$|U|^2$ is the coupling
%strength between the light neutrino and the NHL.
%(The alternative allowed
%decay via a Z boson is not shown.)}
%\label{feynman}
%\end{figure}

%\begin{figure}[tbh]
%\centerline{\psfig{figure=nhl_e.eps,width=0.5\textwidth}}
%\caption{Monte Carlo NHL energy distributions for masses 1.45~GeV and
%0.35~GeV. }
%\label{fig:nhlp}
%\end{figure}

\section*{References}

\begin{references}
\bibitem{grl}  M. Gronau, C.N. Leung, and J.L. Rosner, {\em Phys. Rev.} {\bf D29}, 2539 (1984). \label{bib:grl} 

\bibitem{pdg}  Particle Data Group, C. Caso {\em et al., Eur. Phys. J.} {\bf C3}, 1 (1998). \label%
{bib:pdg} 

\bibitem{malensek} A. Malensek, FNAL-FN-341 (1981) \label{bib:malensek}

\bibitem{turtle}  D.C. Carey, K.L. Brown, F.C. Iselin, SLAC-0246 (1982). \label{bib:turtle}

\bibitem{charm}  R. Ammar {\em et al.,} {\em Phys. Rev. Lett.} {\bf 61}, 2185 (1988); K.~Kodoma {\em et al., Phys. Lett.} {\bf B263}, 573 (1991). (See also S.~Frixione {\em et al., Nucl. Phys.} {\bf B431}, 453 (1994).) \label{bib:charm} 

\bibitem{joe}  J.A. Formaggio {\em et al., Phys. Rev.} {\bf D57}, 7037 (1998) \label%
{bib:joe}. 

\bibitem{tim}  L.M. Johnson, D.W. McKay, and T. Bolton, {\em Phys. Rev.} 
{\bf D56}, 2970 (1997). \label{bib:tim} 

\bibitem{detector}  W. Sakumoto {\em et al., Nucl. Instr. Meth.} {\bf A294}, 
179 (1990). \label{bib:detector} 

\bibitem{lownu}  R. Belusevic and D. Rein, {\em Phys. Rev.} {\bf D38}, 2753 (1988). 
\label{bib:lownu} 

\bibitem{lund}  G. Ingelman {\em et al., Comput. Phys. Commun.}
{\bf 101}, 108 (1997).
\label{bib:lund}      
%\bibitem{lund}  G. Ingelman {\em et al.,} Proc. {\em Physics at HERA, 1992}, (1991). 
%\label{bib:lund}

\bibitem{geant} CERN CN/ASD, GEANT detector description and simulation 
library (1998). \label%
{bib:geant}  

\bibitem{prev_ccfr}  S.R. Mishra {\em et al., Phys. Rev. Lett.} {\bf 59}, 1397 (1987). 
\label{bib:prev_ccfr} 

\bibitem{prev_bebc}  A.M. Cooper-Sarkar {\em et al., Phys. Lett.} {\bf B160}, 207
(1985). \label{bib:prev_bebc} 

\bibitem{prev_charm}  J. Dorenbosch {\em et al., Phys. Lett.} {\bf B166}, 473 (1986). 
\label{bib:prev_charm} 

%\bibitem{prev_kek}  T. Yamazaki {\em et al.,} In Nordkirchen 1984,
%Proceedings of {\em Neutrino Physics and Astrophysics}, published in
%NEUTRINO84, Word Scientific (1985).
%\label{bib:prev_kek}              
\bibitem{prev_kek}  T. Yamazaki {\em et al.,} in {\em Proc. Neutrino 84} (Word Scientific, Singapore, 1985). 
\label{bib:prev_kek} 

\bibitem{prev_lbl}  C.Y. Pang {\em et al., Phys. Rev.} {\bf D8}, 1989 (1973). \label%
{bib:prev_lbl} 
\end{references}
\end{document}